\documentclass[apj]{emulateapj}  
\slugcomment{Astrophysical Journal Letters 803 (2015) L15} %
\usepackage{graphicx,amssymb,amsmath,times}
\usepackage{epsfig,epstopdf}
\usepackage[colorlinks=true,pdfstartview=FitV,linkcolor=blue,citecolor=blue,urlcolor=blue]{hyperref}  
\usepackage{amsmath}
\def\Journal#1#2#3#4{{#4}, {#1}, {#2}, #3} 

\newcommand{\etal}{et al.}

\newcommand{\ApJ}{ApJ}
\newcommand{\AeA}{A\&A}

\newcommand{\PRL}{PRL}
\newcommand{\PRD}{PRD}
\newcommand{\AMS}{\textsf{AMS}} 
\newcommand{\Hyd}{\textsf{H}}
\newcommand{\He}{\textsf{He}}
\newcommand{\Li}{\textsf{Li}}
\newcommand{\Be}{\textsf{Be}}
\newcommand{\B}{\textsf{B}}
\newcommand{\C}{\textsf{C}}
\newcommand{\Oxy}{\textsf{O}}
\newcommand{\Fe}{\textsf{Fe}}
\newcommand{\BC}{\textsf{B/C}}
\newcommand{\eplus}{\ensuremath{e^{+}}}
\newcommand{\eminus}{\ensuremath{e^{-}}}
\newcommand{\epm}{\ensuremath{e^{\pm}}}
\newcommand{\pfrac}{e\ensuremath{^{+}}/(e\ensuremath{^{-}}\,+\,e\ensuremath{^{+}})}

\newcommand{\ie}{\emph{i.e.}}
\newcommand{\eg}{\emph{e.g.}}

\begin{document}

\title{The connection between the positron fraction anomaly and \\ the spectral features in galactic cosmic-ray hadrons}
\shorttitle{Connection between $e^{+}/(e^{+} + e^{-})$ and CR Hadron Spectra}
\shortauthors{N. Tomassetti \& F. Donato}
\author{Nicola Tomassetti$^{1}$ \& Fiorenza Donato$^{2}$}   
\address{$^{1}$\, LPSC, Universit{\'e} Grenoble--Alpes, CNRS/IN2P3, F-38026 Grenoble, France; nicola.tomassetti@lpsc.in2p3.fr;\, }
\address{$^{2}$\, Physics Department, Torino University and INFN, I-10125 Torino, Italy; fiorenza.donato@to.infn.it}

\begin{abstract} 
%
Recent data on Galactic cosmic-ray (CR) leptons and hadrons gave rise to two exciting problems: on the lepton side, 
the origin of the rise of the CR positron fraction \pfrac{} at $\sim$\,10 -- 300 GeV of energy;  
on the hadron side, the nature of the spectral hardening observed in CR protons and nuclei 
at $\sim$\,TeV energies. The lepton anomaly indicates the existence of a nearby \epm{} source.
It has been proposed that high-energy positrons can be produced inside nearby supernova remnants (SNRs) via 
interactions of CR hadrons with the ambient medium. A distinctive prediction of this mechanism is a high-energy 
rise of the boron-to-carbon ratio, which has not been observed.
It also requires \textit{old} SNRs at work (with ineffective magnetic field amplification and slow shock speed),
that cannot account for the CR hadronic spectra observed up to the knee energies ($\sim$\,5\,PeV).
We propose a new picture where, in addition to such a nearby CR accelerator, the high-energy spectrum of CR hadrons 
is provided by the large-scale population of SNRs, on average younger, that can efficiently accelerate CRs up to the knee. 
Under this scenario, the spectral hardening of CR hadrons can be naturally interpreted as the transition between the two components.
As we will show, our two-component model breaks the connection between the positron fraction and the boron-to-carbon 
ratio, which is now predicted to decrease with energy in accordance with the data.
Forthcoming data from \AMS{} will be crucial for testing this model.
\end{abstract}
\keywords{cosmic rays --- acceleration of particles --- ISM: supernova remnants --- nuclear reactions, nucleosynthesis, abundances}

\section{Introduction}    
\label{Sec::Introduction} 
%
The \AMS{} collaboration has recently confirmed with high precision the CR positron fraction anomaly previously 
observed by PAMELA \citep{Adriani2009,Aguilar2013}. The data show a rise of the fraction up at energies
$\sim$\,10 -- 300\,GeV, followed by a possible plateau at higher energies \citep{Accardo2014},
which cannot be described by conventional models of \eplus{} production by collisions of CR hadrons with the interstellar matter (ISM).
In these models, the primary CRs (\eg, \eminus, \Hyd, \He, \C, or \Oxy{}) are 
accelerated by supernova remnants (SNRs) via diffusive shock acceleration (DSA) mechanisms up to $\sim$\,PeV energies 
to power-law spectra $S \propto E^{-\nu}$. Their subsequent propagation is described by means of an energy-dependent 
confinement time $\tau_{\rm esc}\propto E^{-\delta}$ (or diffusion coefficient $K\propto E^{\delta}$),
and their collisions with the ISM give rise to \textit{secondary} CRs such as \Li, \Be, \B, \eplus, or $\bar{p}$ \citep{Strong2007}. 
At $E \gg$\,GeV, this picture predicts power-law spectra for primary nuclei, $N_{p}\sim E^{-\nu - \delta}$,
and decreasing secondary-to-primary ratios, \eg, the boron-to-carbon: \BC\,$\sim E^{-\delta}$, 
where $\delta\sim$\,0.3--0.7, $\nu\sim$\,2.0--2.4, and $\nu+\delta\approx$\,2.7. 
The high-energy spectrum of CR leptons is further steepened by radiative losses, with characteristic time-scale 
$\tau^{\rm rad}(E) \propto E^{-1}$. Given all the positrons are from secondary origin, the positron fraction is expected 
to decrease similarly to other secondary-to-primary ratios, in remarkable contrast with the observations.
Explanations of the rise may include either dark-matter particles annihilation or decay, or nearby astrophysical 
sources \citep{Serpico2011}.  Among the second class, it has been proposed that high energy positrons may be produced 
through hadronic interactions  of CR protons undergoing DSA inside \textit{old} SNRs \citep{Blasi2009}. 
Yet, if secondary positrons are produced and accelerated by this mechanism, other secondary species 
(such as \B{} nuclei or $\bar{p}$) will also be produced from CR nuclei interacting with the gas. 
As shown by \citet{MertschSarkar2009}, this mechanism leads to an increase of the \BC{} ratio at $\gtrsim$\,100\,GeV per nucleon.
However the current \BC{} ratio data decrease with energy, indicating that the \textit{old} SNR scenario (hereafter OSNR) 
should be ruled out \citep{CholisHooper2014}.

On the other hand, the spectra of primary CR elements have been measured up to $\sim$\,PeV energies and beyond.
Recent data on CR protons and nuclei revealed a remarkable spectral hardening at $\sim$\,TeV energies which 
stimulated great interest \citep{Blasi2013}.
According to the PAMELA data on \Hyd{} and \He{} \citep{Adriani2011}, the change of slope is located 
at $R\approx$\,230\,GV of rigidity (\ie, momentum-to-charge ratio) with a very sharp transition, 
which is not seen by other experiments. 
While the sharpness of such breaks is under debate, the CR spectral hardening at $E\gtrsim$\,1\,TeV per 
nucleon is established by several experiments such as CREAM and ATIC \citep{Panov2009,Yoon2010}.
The proposed explanations invoke acceleration mechanisms \citep{Ptuskin2013}, diffusion effects \citep{Tomassetti2012,Blasi2012},
or the superposition of local and distant sources \citep{Bernard2013,Thoudam2013,Vladimirov2012}. 

In this Letter, we argue that the OSNR scenario is incomplete in order to account for the observations of CR hadron spectra.
Whether the OSNR represents a single source or a population of sources with the characteristics required for producing 
secondary $e^{\pm}$ (\ie, low shock speed and damped magnetic fields), it cannot provide the flux of CR hadrons observed 
in the $\sim$\,TeV -- PeV energy region, so that an additional high-energy component of CR accelerators is needed. 
We propose a two-component SNR scenario where the high-energy part of the CR flux is provided by a Galactic ensemble of SNRs, 
hereafter GSNR, that are on average younger and more efficient
when accelerating primary hadrons at high energies (but are unable to accelerate secondaries). 
A key consideration is that the local flux of light CR nuclei depends on the large-scale structure of the Galaxy,
reflecting the contribution of a large population of SNRs and their histories \citep{Taillet2003}.
For $E\gtrsim$\,10\,GeV, their escape rate, $\tau_{\rm esc}^{-1}$, is generally larger than their spallation rate 
in the ISM $\tilde{\Gamma}^{\rm inel}$ so that their propagation is limited only by the size of the diffusion region.
In contrast, light leptons are subjected to radiative cooling, due to synchrotron radiation and inverse Compton scattering,
which limits significantly the range that they can travel before reaching the solar system \citep{Delahaye2010}.
Their characteristic distance is approximately $\lambda \sim \sqrt{\tau^{\rm rad} K} \propto E^{(\delta-1)/2}$, confined to 
$\sim$\,1\,kpc at energy above $\sim$\,100\,GeV (depending on the propagation parameters), which would legitimate the OSNR approach.
Thus, while the observed \epm{} can be largely produced by only one or few nearby sources, the total spectrum of CR protons 
and nuclei may well result as the sum two SNR components: the nearby OSNR component, which would dominate the flux below 
$\sim$\,100\,GeV, and the GSNR ensemble, which would provide the high-energy flux up to the knee.
As we will show, a two-component model gives a good description of the primary CR spectral hardening and it has an impact 
on the spectral shape of the \BC{} ratio, which is now determined as the superposition of several contributions.

\section{Calculations}     
\label{Sec::Calculations}  
%
The spectrum of CRs accelerated in SNRs is computed analytically using the linear
DSA theory and including the secondary production terms due to hadronic interactions. 
We follow our calculation scheme in \citet{TomassettiDonato2012}, which has been proven
to be equivalent to that of \citet{MertschSarkar2009,MertschSarkar2014}.
In the shock rest-frame ($x = 0$), the upstream plasma flows in from $x < 0$ with speed $u_{1}$ (density $n_{1}$) 
and the downstream plasma flows out to $x > 0$ with speed $u_{2}$ (density $n_{2}$). The compression ratio 
is $r=u_{1}/u_{2} = n_{2}/n_{1}$. For a nucleus with charge $Z$ and mass number $A$, the DSA equation reads:
\begin{equation} \label{Eq::DiffusionDSA}
 u \frac{\partial f}{\partial x} = D \frac{\partial^{2}
 f}{\partial x^{2}} + 
 \frac{1}{3}\frac{du}{dx}p\frac{\partial f}{\partial p} 
 -\Gamma^{\rm inel}{f}  + Q \,,
\end{equation}
where $f$ is the phase space density, $D(p)$ is the diffusion coefficient near the SNR shock, 
$u$ is the fluid velocity and $\Gamma^{\rm inel} = c \beta n \sigma^{\rm inel}$ is the total destruction 
rate for fragmentation with the interaction cross-section $\sigma^{\rm inel}$.
The ambient density $n$ is assumed to be composed by 90\,\% \Hyd{} and 10\,\% \He, similarly to the average ISM, in both sides of the shock. 
The source term is represented by $Q$. For primary nuclei 
$Q^{\rm pri} = Y \delta(x) \delta(p-p^{\rm inj})$, \ie,
the ambient particle injection occurs immediately upstream the shock at momentum $p^{\rm inj}\equiv Z R^{\rm inj}$, where 
$R^{\rm inj}\equiv 1\,$GV for all species. The $Y$-constants are abundance factors, determined from the data. 
The source term for secondary fragments produced by spallation of heavier ($k$--labeled) nuclei is of the type 
$Q^{\rm sec}=\sum_{k}\Gamma_{k}^{\rm frag}f_{k}$, where the partial rates of fragmentation are $\Gamma_{k}^{\rm frac}=c\beta n \sigma_{k}^{\rm frag}$.
Using the subscript $i=1$ ($i=2$) to indicate the quantities in the upstream (downstream) 
region, the downstream solution of each nucleus can be expressed as:
\begin{equation}\label{Eq::DownStreamSecondary} 
f_{2}(x,p) = f_{0} +  \frac{r x }{u_{2}} \left( Q_{0} - \Gamma^{\rm inel}_{1} f_{0} \right) \,,
\end{equation}
where the subscript $i=0$ indicates the quantities evaluated at the shock ($x=0$), and $f_{0}(p)$
is given by:
\begin{equation}\label{Eq::FullSolutionAtShockFront} 
  \begin{aligned} 
    f_{0}(p) =  \alpha \int_{0}^{p} \left( \frac{p'}{p} \right)^{\alpha} Q^{\rm pri}(p')  e^{-\chi(p,p')} \frac{dp'}{p'}\\ 
   + \alpha \int_{0}^{p} \left( \frac{p'}{p} \right)^{\alpha} \frac{Q^{\rm sec}_{1}(p') D}{u^{2}_{1}}\left( 1 + r^{2} \right) e^{-\chi(p,p')} \frac{dp'}{p'} \,,
\end{aligned}
\end{equation}
with $\alpha = 3r/(r-1)$ and $\chi \approx  \alpha(\Gamma^{\rm inel}_{1}/\Gamma^{\rm acc}) [ D(p) - D(p') ]$,
where $\Gamma^{\rm acc}$ is the acceleration rate.
The first term of Eq.\,\ref{Eq::FullSolutionAtShockFront} gives the acceleration spectrum of primary particles at 
the shock, of the form $f^{\rm pri}_{0}\sim p^{-\alpha} e^{-\chi}$. 
The second term describes the production and acceleration of CR fragments and it is coupled with the equations of heavier nuclei.
The amount of secondary nuclei production depends on the SNR properties via $n_{1} D{u_{1}^{-2}}$.
Their spectrum is of the form $f^{\rm sec}_{0}\sim f^{\rm pri}_{0}D(p)e^{-\chi} \propto p^{-\alpha+1}e^{-\chi}$, \ie, 
$D$ times harder than the primary source spectrum.
For having an efficient acceleration of all particles, the condition $\Gamma^{\rm inel}_{1} \ll \Gamma^{\rm acc}$ must be fulfilled 
in the whole momentum (or rigidity) range considered \citep{MertschSarkar2009}. For Bohm-type diffusion ($D = \frac{R}{3B}$), 
this condition can be also expressed as $R\ll R^{\rm c} \sim \frac{3Bu_{1}^{2}}{20 c \Gamma^{\rm inel}}$.
At $R \gtrsim R^{\rm c}$ one has $\Gamma^{\rm inel}_{1} \gtrsim \Gamma^{\rm acc}$, \ie, destructive interactions dominate over acceleration.
In case of no interactions ($Q=0$ and $\Gamma^{\rm inel/frag}=0$) the usual DSA solution $f\propto p^{-\alpha}$ 
is recovered for primary nuclei, while the secondary CR production vanishes.
The total CR flux produced by the SNR is obtained by integration in its volume: 
\begin{equation}\label{Eq::SNRVolumeIntegral} 
S^{\rm snr}(p) = 4\pi p^{2} e^{-p/p^{\rm max}} \int_{0}^{\tau^{\rm snr}u_{2}} 4 \pi x^{2} f_{2,j}(x,p) dx
\end{equation}
The exponential cut-off $e^{-p/p^{\rm max}}$ is used to explicitly account for the maximum momentum attained 
by the SNR, due to the finite time of the DSA process $\tau^{\rm snr}$, and it is assumed to occur at the same 
rigidity for all CRs in the accelerator: $R^{\rm max}\equiv p^{\rm max}/Z$.
$R^{\rm max}$ can be roughly estimated from equating $\Gamma^{\rm acc}$ with $1/\tau^{\rm snr}$, 
which gives $\hat{R}^{\rm max} \sim  0.2 B u_{1}^{2}c^{-1}\tau^{\rm snr}$,
though it is usually left as free parameter determined from the data \citep{Kachelriess2011,Serpico2011}.
We stress that the steady-state description given here is an effective simplification of a 
more complex physical problem where the shock properties evolve with time. 

For the Galactic propagation, we use an analytical approach of CR diffusion and nuclear interactions,
where the effects of energy changes are disregarded \citep{Maurin2001}. 
The Galaxy is modeled as a disc of half-thickness $h$ containing the ISM gas (number density $n_{\rm ism}$) 
and the CR sources. The disc is surrounded by a cylindric diffusive halo of half-thickness $L$ and zero matter density. 
For each CR nucleus, the transport equation reads:
\begin{equation}\label{Eq::Transport1D}
\frac{\partial N}{\partial t} = K \frac{ \partial^{2} N}{\partial z^{2}} -2h \delta(z) \tilde{\Gamma}^{\rm inel} N + 2h\delta(z) S\,,
\end{equation}
where $N(z)$ is its number density as function of the $z$-coordinate, $K$ is the Galactic diffusion 
coefficient and $\tilde{\Gamma}^{\rm inel} = \beta c n_{\rm ism} \sigma^{\rm inel}$ is the destruction rate 
in the ISM at velocity $\beta c$ and cross section $\sigma^{\rm inel}$. 
The source term $S$ includes the SNR contributions, $S^{\rm snr}$, and term for secondary production in the ISM 
from spallation of heavier ($k$) nuclei: $S^{\rm ism}= \sum_{k} \tilde{\Gamma}^{\rm frag}_{k}N_{k}$.
The diffusion coefficient is taken as $K(R) = \beta K_{0}(R/R_{0})^{\delta}$, spatially homogeneous,
where $K_{0}$ expresses its normalization at $R_{0}\equiv$\,4\,GV.
We solve Eq.\,\ref{Eq::Transport1D} for all nuclei (from \Fe{} to \Hyd) after assuming stationarity ($\partial N/\partial t =0$), 
boundary conditions ($N(\pm L)\equiv 0$), and continuity condition across the disc.
The differential fluxes at Earth are given by $\phi(E) = \frac{\beta c}{4 \pi}N_{0}(E)$, where $N_{0}$, evaluated 
at $z=0$, is of the type $N_{0}\approx S \left( \frac{K}{hL} + \tilde{\Gamma}^{\rm inel} \right)^{-1}$.
The quantities $N$, $K$, $S$ and $\tilde{\Gamma}$ depend on energy or rigidity too. 
To account for the solar modulation, we employ the \textit{force-field} approximation \citep{Gleeson1968}
using the parameter $\Phi=500$\,MV for a medium-level modulation strength.

\section{Results and Discussion}    
\label{Sec::TwoComponent}           
%
There are many parameters that determine the OSNR source spectra. 
We follow the benchmark model of \citet{MertschSarkar2014}, that provides good fits to the \AMS{} 
leptonic data, assuming that the bulk of the \epm{} flux is produced by this type of OSNRs.
\setlength{\tabcolsep}{0.036in} 
\begin{table}[!h]
\caption{ 
  Source and transport parameter sets.
\label{Tab::Parameters}}
\centering
\begin{tabular}{ccccc} \hline\hline
\multicolumn{2}{c}{ OSNR parameters } & {\hspace{1.5cm}} & \multicolumn{2}{c}{Propagation parameters} \\
\hline
$u_{1}$               &  5$\times$10$^{7}$\,cm\,s$^{-1}$     &     &   $K_0$  & 0.1\,kpc$^{2}$\,Myr$^{-1}$ \\
$B$ / $\kappa_{B}$        &  1\,$\mu$G / 16     &     &   $\delta$    &  0.50 \\
$\alpha_{H}$/$\alpha_{Z>1}$     &  4.65/4.55    &     &    $L$   &  5\,kpc\\
$n_{1}$                &  2\,cm$^{-3}$     &     &$h$    & 0.1\,kpc     \\
$R^{\rm max}$           &   1\,TV     &     &       $n_{\rm ism}$    &   1\,cm$^{-3}$ \\
$\tau^{\rm snr}$        &   50\,kyr    &     &  $\Phi$    &  0.5\,GV \\ 
\hline
\end{tabular}
\end{table}
All relevant parameters are listed in Tab.\,\ref{Tab::Parameters}. In particular we adopt $B=$\,1\,$\mu$G, $R^{\rm max}=$\,1\,TV,  
$\kappa_{B}=$\,16, and $u_{1}=$\,5$\times$\,10$^{7}$\,cm\,s$^{-1}$, where $\kappa_{B}$ parametrizes the deviation of $D(p)$ 
from the Bohm value due to magnetic damping. These values are typical for SNRs at their late evolutionary stages. 
\begin{figure}
\begin{center}
\epsscale{0.95}
\plotone{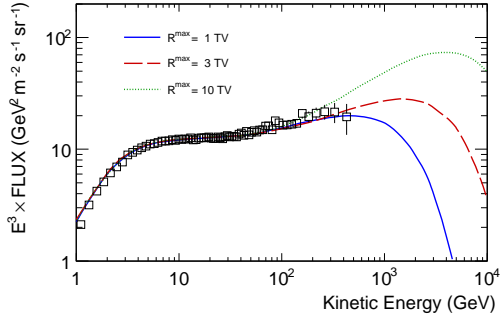}
\figcaption{ 
  Energy spectrum of CR positrons multiplied by $E^{3}$.
  The three models of \citet{MertschSarkar2014} (lines)
  are compared with the new data from \AMS{} \citep{Aguilar2014}.
  \label{Fig::ccPositronFlux}
}
\end{center}
\end{figure}
The authors in \citet{MertschSarkar2014} considered also scenarios with higher values of $R^{\rm max}$, fixed at  
3 TV and 10 TV, which can in principle discriminated with $e^{+}$ data at higher energy. 
In Fig.\,\ref{Fig::ccPositronFlux} we compare these predictions with the new high-energy data released by \AMS{} \citep{Accardo2014,Aguilar2014}. 
The data suggest that models with high $R^{\rm max}$ ($\sim$\,10\,TV or higher) are disfavored.
We also note that the value $R^{\rm max}$=\,1\,TV is consistent with the naive estimate made from equating $\Gamma^{\rm acc}$ with  $1/\tau^{\rm snr}$.
At this point it is clear that a pure OSNR scenario, which describes well the $\sim$\,GV - TV observations, cannot account for the 
CR hadronic flux observed at $\sim$\,TV - PV rigidities. This is also the rigidity region where the spectra are found to be harder. 
\begin{figure}
\begin{center}
\epsscale{1.0}
\plotone{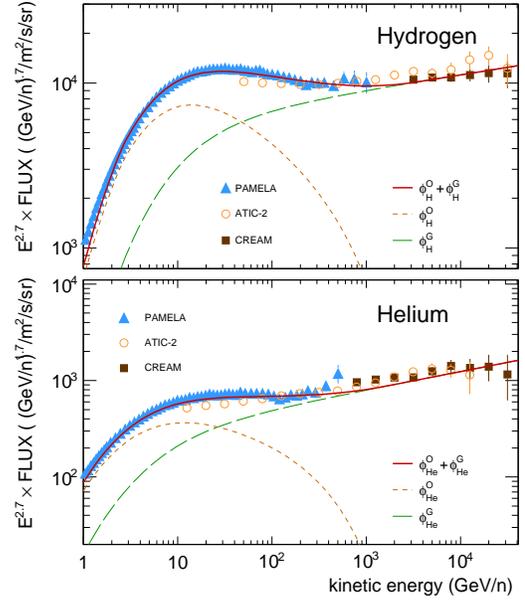}
\figcaption{ 
  Energy spectra of \Hyd{} (top) and \He{} (bottom) multiplied by $E^{2.7}$.
  The solid lines indicate the model calculations. The contribution arising from
  OSNR (short-dashed lines) and from GSNR (long-dashed lines) are shown.
  The data are from PAMELA \citep{Adriani2011}, ATIC-2 \citep{Panov2009} and CREAM \citep{Yoon2010}. 
  \label{Fig::ccHadronSpectra}
}
\end{center}
\end{figure}
This consideration motivated us to introduce a second component for the CR hadron spectra at high energies, \ie, 
the GSNR component, representing the large-scale population of distant SNRs.
Typical parameters for GSNRs with strong shock and amplified magnetic fields are $u_{1}\sim$\,10$^{9}$\,cm\,s$^{-1}$,
$B/\kappa_{B}\sim$\,100\,$\mu$G, and $R^{\rm max}\sim$\,5\,PV.
It is easy to see that, from these values, GSNRs are unable to produce and accelerate secondary $e^{\pm}$ or Li-Be-B. 
Furthermore, the resulting CR spectra are totally insensitive to their exact values (and to the type of diffusion) 
so that the only relevant GSNR parameters are the source spectral indices.
For both components, OSNR and GSNR, the slope $\alpha$ and theirnor normalization are chosen to match the data 
on primary spectra after propagation. The source parameter $\alpha$ is degenerated with the transport parameter 
$\delta$, but the latter can been tested against the \BC{} ratio.
As in \citet{MertschSarkar2014} and related works, for  $Z=1$ we use a source spectral index steeper by 0.1 
compared to that of heavier nuclei. This is a known issue, possibly ascribed to an $A/Z$--dependent 
injection efficiency in SNR shocks \citep{Malkov2012}.
The relative source abundances are those adopted from previous studies \citep{Tomassetti2012,TomassettiDonato2012} 
and we use the same values for the two SNR components. The contributions of the two components,
determined from the data, are taken as 85\,\% for the OSNR and 15\,\% for the GSNR flux at 1\,GeV/n, for all elements.
Leptonic spectra from GSNR are expected not to contribute significantly to the high-energy flux, which is 
the case if these sources are placed at distances $d\gtrsim$\,kpc \citep{Delahaye2010}. 
The data at $\gtrsim$\,TeV energies require the GSNR spectra to be harder than those from the OSNR: we adopt 
$\alpha_{H}=$\,4.1 and $\alpha_{Z>1}=$\,4.0. This is in fact encouraging, because the basic DSA predictions, 
supported by $\gamma$--ray observations of young SNRs, favor $\alpha \sim$\,4.0 -- 4.2 \citep{Blasi2013}. 
With this setup, in Fig.\,\ref{Fig::ccHadronSpectra} we plot the model predictions for \Hyd{} and \He{} fluxes 
as function of kinetic energy per nucleon. The two components of the flux are shown, \ie, split 
as $\phi_{H}=\phi_{H}^{O} + \phi_{H}^{G}$, and $\phi_{He}=\phi_{He}^{O} + \phi_{He}^{G}$. 
The data are well described by the model, which interprets the TeV spectral hardening in terms of a smooth transition 
between the OSNR and the GSNR component, but the fine structures of the PAMELA data cannot be recovered.
These sharp breaks seem also to be in contrast with the preliminary results of \AMS. 
We eagerly await the \AMS{} final results on \Hyd{} and \He{} spectra at high energy,
that will hopefully clarify how and where the spectral transition take places.

\begin{figure}
\begin{center}
\epsscale{1.0}
\plotone{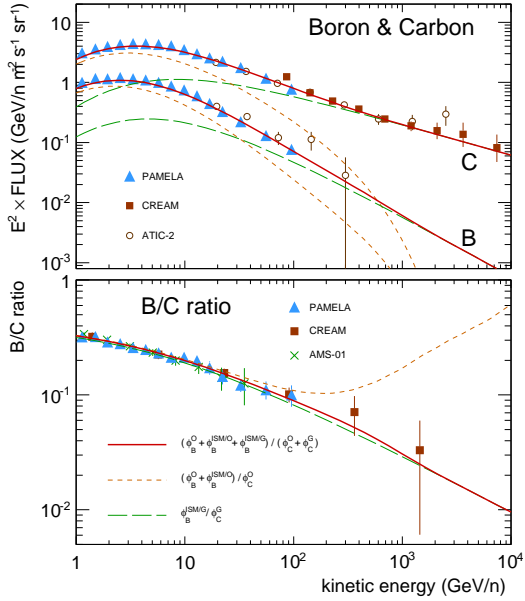}
\figcaption{ 
  Top: CR energy spectra of \B{} and \C{} multiplied by $E^{2}$.
  Bottom: \BC{} ratio.
  The lines indicate the model calculations. In the top panel, the \B{} components are
  for $\phi_{\rm B}^{\rm O}+\phi_{\rm B}^{\rm ISM/O}$ (short-dashed line), 
  and $\phi_{\rm B}^{\rm ISM/G}$ (long-dashed line) and their sum (solid line). 
  The \C{} components are for $\phi_{\rm C}^{\rm O}$ (short-dashed line)
  and $\phi_{\rm C}^{\rm G}$ (long-dashed line) and their sum (solid line).
  The data are from and PAMELA \citep{Adriani2014}, CREAM \citep{Yoon2010}, \AMS-01 \citep{Aguilar2010},
  and ATIC-2 \citep{Panov2009}. 
  \label{Fig::ccNucleiSpectra}
}
\end{center}
\end{figure}

We now come to the \BC{} ratio. Figure\,\ref{Fig::ccNucleiSpectra} shows the \B{} and \C{} fluxes (top) and the \BC{} 
ratio (bottom) compared with recent data, including the new data from PAMELA \citep{Adriani2014}.
The carbon flux, mostly primary, is also of the type $\phi_{\rm C}\approx \phi_{\rm C}^{\rm O} + \phi_{\rm C}^{\rm G}$.
It also experience a spectral hardening that is well reproduced by the model. 
The \B{} spectrum, entirely secondary, can be ideally split into $\phi_{\rm B}=\phi_{\rm B}^{\rm O} + \phi_{\rm B}^{\rm ISM}$,
where the first component is the one produced inside the OSNR, and the second arises in the ISM via 
collisions of heavier nuclei such as \C, \Oxy, or \Fe. 
Thus, the ISM component $\phi_{\rm B}^{\rm ISM}$ can be split again into $\phi_{\rm B}^{\rm ISM/O}$ 
(produced by collisions of OSNR-emitted nuclei) and $\phi_{\rm B}^{\rm ISM/G}$ (by GSNR-emitted nuclei).
In previous OSNR--related works such as \citet{CholisHooper2014}, the \BC{} ratio is always meant 
as $(\phi_{\rm B}^{\rm O}+\phi_{\rm B}^{\rm ISM/O})/\phi_{\rm C}^{\rm O}$. 
Remarkably, while the ratio $(\phi_{\rm B}^{\rm O}+\phi_{\rm B}^{\rm ISM/O})/\phi_{\rm C}^{\rm O}$
starts rising at $E \gtrsim$\,100\,GeV per nucleon (as expected by a pure OSNR scenario), 
the ratio of the two-component model $\phi_{\rm B}/\phi_{\rm C}$ decreases with energy in good agreement with the data. 
The trend of our \BC{} ratio is similar to the one from conventional propagation models (long-dashed line) where only 
GNRS component is considered.
This effect can be understood from the top panel of the figure: at $\sim$\,100\,GeV/nucleon, when the ONSR component of \B{}
would become relevant enough to provide a signature (\ie, $\phi_{\rm B}^{\rm O} \gtrsim \phi_{\rm B}^{\rm ISM/O}$), 
the total fluxes of \B{} and \C{} are both dominated by the GSNR components, $\phi_{\rm B}^{\rm ISM/G}$ and $\phi_{\rm C}^{\rm G}$ respectively.
Thus, in our two-component scenario, the \BC{} ratio does not constrain secondary production in SNR, as the presence 
of the GSRN component breaks the connection between the positron fraction and the secondary-to-primary nuclear ratios.
We have tested the calculation using different values of $R^{\rm max}$. But due to a degeneration with the 
OSNR parameter $\alpha$, the parameter $R^{\rm max}$ cannot be constrained much within the precision of the data.
Our results for the \BC{} ratio seem to be quite robust: once accounting for the GSNR component, it decreases with 
energy as $\sim E^{-\delta}$  (giving only little wiggles around the energy $E \approx R^{\rm max}/2$, see Fig.\,\ref{Fig::ccNucleiSpectra}).
For $R^{\rm max}\sim$\,1\,TV or less, the spectra of \Hyd{} and \He{} experience
deviations from the power-law behavior below the TeV region that should be measurable by \AMS.

\section{Conclusions}            
\label{Sec::Conclusions}         
%
Our work is motivated by two outstanding problems in CR physics: the origin of the rise in 
the positron fraction at $E\gtrsim$\,10\,GeV, and the  nature of the $\sim$\,TeV spectral hardening in CR hadrons.
We revisited the ONSR scenario, proposed for the positron fraction anomaly, in order to account for the high-energy 
observations of CR hadron and nuclei. In OSNR models, secondary particles such as positrons and light-nuclei 
are produced and accelerated inside SNRs via hadronic interactions. 
In order to be able to match the $e^{\pm}$ data, these old-SNRs must have particular properties in terms of environmental 
parameters (such as strongly damped magnetization, or relatively high gas density) and predicts features in the B/C ratio that are not observed.
In this Letter, we have argued that the OSNR scenario is incomplete for explaining the flux of CR hadrons at $\sim$\,TeV -- PeV energies. 
The OSNR can account for the leptonic flux and for the GeV-TeV production of CR hadrons, but the flux at higher energies has 
to be provided by a population of distant and young sources that are able to efficiently accelerate CRs up to the knee. 
These sources are unable to produce secondary CRs (due to magnetic field amplification) and they do
not contribute significantly to the high-energy leptonic flux (due to the large distance).
Within this picture the spectral hardening of CR hadrons is interpreted as a signature of the transition between the 
local OSNR component and the Galactic ensemble. The spectra of all primary nuclei are predicted to harden.
Taking into account the contribution of the two populations, we found that the predicted \BC{} ratio show no prominent signatures:
it decreases with energy in accordance with the existing data. In conclusion, this generalized scenario may explains the absence 
of signatures in the \BC{} ratio while accounting for the observed signatures in primary CR hadrons.
A quantitative inspection will be done with a proper modeling of leptonic and hadronic spectra arising from a realistic time-space 
distribution of their sources. It will be crucial, for achieving this goal, to have precision data on CR protons and nuclei in the 
energy region where the spectral transition takes place.   
\\
\\
\footnotesize{
N.T. acknowledges the support of the Labex grant \textsf{ENIGMASS}.
The work of F.D. is supported by the  {\sl Strategic Research Grant:
Origin and Detection of Galactic and Extragalactic Cosmic Rays} funded by
Torino University and Compagnia di San Paolo and by the research grant
{\sl Theoretical Astroparticle Physics} N. 2012CPPYP7 under the program 
\textsf{PRIN-2012} funded by the MIUR, Italy. 
}


\end{document}